# SYMBOLIC COMPUTATION OF RECURSION OPERATORS FOR NONLINEAR DIFFERENTIAL-DIFFERENCE EQUATIONS

Ünal Göktaş[1*] and Willy Hereman[2]
[1]Department of Computer Engineering, Turgut Özal University
Keçiören, Ankara 06010, Turkey.
ugoktas@turgutozal.edu.tr
[2]Department of Mathematical and Computer Sciences, Colorado School of Mines
Golden, Colorado 80401-1887, U.S.A.
whereman@mines.edu

**Abstract-** An algorithm for the symbolic computation of recursion operators for systems of nonlinear differential-difference equations (DDEs) is presented. Recursion operators allow one to generate an infinite sequence of generalized symmetries. The existence of a recursion operator therefore guarantees the complete integrability of the DDE. The algorithm is based in part on the concept of dilation invariance and uses our earlier algorithms for the symbolic computation of conservation laws and generalized symmetries.

The algorithm has been applied to a number of well-known DDEs, including the Kac-van Moerbeke (Volterra), Toda, and Ablowitz-Ladik lattices, for which recursion operators are shown. The algorithm has been implemented in *Mathematica*, a leading computer algebra system. The package **DDERecursionOperator.m** is briefly discussed.

**Keywords-** Conservation Law, Generalized Symmetry, Recursion Operator, Nonlinear Differential-Difference Equation

## 1. INTRODUCTION

A number of interesting problems can be modeled with nonlinear differential-difference equations (DDEs) [1]-[3], including particle vibrations in lattices, currents in electrical networks, and pulses in biological chains. Nonlinear DDEs also play a role in queuing problems and discretizations in solid state and quantum physics, and arise in the numerical solution of nonlinear PDEs.

The study of complete integrability of nonlinear DDEs largely parallels that of nonlinear partial differential equations (PDEs) [4]-[7]. Indeed, as in the continuous case, the existence of large numbers of generalized (higher-order) symmetries and conserved densities is a good indicator for complete integrability. Albeit useful, such predictors do not provide proof of complete integrability. Based on the first few densities and symmetries, quite often one can explicitly construct a recursion operator which maps higher-order symmetries of the equation into new higher-order symmetries. The existence of a recursion operator, which allows one to generate an infinite set of such symmetries step-by-step, then confirms complete integrability.

There is a vast body of work on the complete integrability of DDEs. Consult, e.g., [5, 8] for additional references. In this article we describe an algorithm to symboli-cally compute recursion operators for DDEs. This algorithm builds on our related work for PDEs and DDEs [9]-[11] and work by Oevel *et al* [12] and Zhang *et al* [13].

In contrast to the general symmetry approach in [5], our algorithms rely on specific assumptions. For example, we use the dilation invariance of DDEs in the construction of densities, higher-order symmetries, and recursion operators. At the cost of generality, our algorithms can be implemented in major computer algebra systems.

Our *Mathematica* package **InvariantsSymmetries.m** [14] computes densities and generalized symmetries, and therefore aids in automated testing of complete integrability of semi-discrete lattices. Our new *Mathematica* package **DDERecursionOperator.m** [15] automates the required computations for a recursion operator.

The paper is organized as follows. In Section 2, we present key definitions, necessary tools, and prototypical examples, namely the Kac-van Moerbeke (KvM) [16] and Toda [17, 18] lattices. An algorithm for the computation of recursion operators is outlined in Section 3. Usage of our package is demonstrated on an example in Section 4. Section 5 covers two additional examples, namely the Ablowitz-Ladik (AL) [19] and RelativisticToda (RT) [20] lattices. Concluding remarks about the scope and limitations of the algorithm are given in Section 6.

## 2. KEY DEFINITIONS

### 2.1. Definition

A nonlinear DDE is an equation of the form

$$\dot{\mathbf{u}}_n = \mathbf{F}(\ldots, u_{n-1}, u_n, u_{n-1}, \ldots), \tag{1}$$

where $\mathbf{u}_n$ and $\mathbf{F}$ are vector-valued functions with $N$ components. The subscript $n$ corresponds to the label of the discretized space variable; the dot denotes differentiation with respect to the continuous time variable $t$. Throughout the paper, for simplicity we denote the components of $\mathbf{u}_n$ by $(u_n, v_n, w_n, \ldots)$ and write $\mathbf{F}(\mathbf{u}_n)$, although $\mathbf{F}$ typically depends on $\mathbf{u}_n$ and a finite number of its forward and backward shifts. We assume that $\mathbf{F}$ is polynomial with constant coefficients. No restrictions are imposed on the shifts or the degree of nonlinearity in $\mathbf{F}$.

### 2.2. Example

The Kac-van Moerbeke (KvM) lattice [16], also known as the Volterra lattice,

$$\dot{u}_n = u_n(u_{n+1} - u_{n-1}), \tag{2}$$

arises in the study of Langmuir oscillations in plasmas, population dynamics, etc.

### 2.3. Example

One of the earliest and most famous examples of completely integrable DDEs is the Toda lattice [17,18],

$$\ddot{y}_n = \exp(y_{n-1} - y_n) - \exp(y_n - y_{n+1}), \tag{3}$$

where $y_n$ is the displacement from equilibrium of the $n$th particle with unit mass under an exponential decaying interaction force between nearest neighbors. With the change of variables, $u_n = \dot{y}_n$, $v_n = \exp(y_n - y_{n+1})$, due to Flaschka [21], lattice (3) can be written in polynomial form [22]

$$\dot{u}_n = v_{n-1} - v_n, \quad \dot{v}_n = v_n(u_n - u_{n+1}). \tag{4}$$

**2.4. Definition**

A DDE is said to be dilation invariant if it is invariant under a scaling (dilation) symmetry.

**2.5. Example**

Lattice (2) is invariant under $(t, u_n) \to (\lambda^{-1}t,\ \lambda\, u_n)$, where $\lambda$ is an arbitrary scaling parameter.

**2.6. Example**

Equation (4) is invariant under the scaling symmetry

$$(t, u_n, v_n) \to (\lambda^{-1}t,\ \lambda\, u_n, \lambda^2\ v_n), \tag{5}$$

where $\lambda$ is an arbitrary scaling parameter.

**2.7. Definition**

We define the weight, $w$, of a variable as the exponent in the scaling parameter $\lambda$ which multiplies the variable. As a result of this definition, $t$ is always replaced by $\frac{t}{\lambda}$ and $w(\mathrm{d}/\mathrm{dt}) = w(\mathrm{D}_t) = 1.$ In view of (5), we have $w(u_n) = 1,$ and $w(v_n) = 2$ for the Toda lattice.

Weights of dependent variables are nonnegative, integer or rational numbers, and independent of $n$. For example, $w(u_{n-1}) = w(u_n) = w(u_{n+1}),$ etc.

**2.8. Definition**

The rank of a monomial is defined as the total weight of the monomial. An expression is uniform in rank if all of its terms have the same rank.

**2.9. Example**

In the first equation of (4), all the monomials have rank 2; in the second equation all the monomials have rank 3. Conversely, requiring uniformity in rank for each equation in (4) allows one to compute the weights of the dependent variables (and thus the scaling symmetry) with elementary linear algebra. Indeed,

$$w(u_n) + 1 = w(v_n), \quad w(v_n) + 1 = w(u_n) + w(v_n), \tag{6}$$

yields

$$w(u_n) = 1, \quad w(v_n) = 2, \tag{7}$$

which is consistent with (5).

Dilation symmetries, which are Lie-point symmetries, are common to many lattice equations. Polynomial DDEs that do not admit a dilation symmetry can be made scaling invariant by extending the set of dependent variables with auxiliary parameters with appropriate scales.

**2.10. Definition**

A scalar function $\rho_n(\mathbf{u}_n)$ is a conserved density of (1) if there exists a scalar function $J_n(\mathbf{u}_n)$, called the associated flux, such that [23]

$$\mathrm{D}_t \rho_n + \Delta\, J_n = 0 \tag{8}$$

is satisfied on the solutions of (1).

In (8), we used the (forward) difference operator,

$$\Delta\, J_n = (\mathrm{D} - \mathrm{I})\, J_n = J_{n+1} - J_n, \tag{9}$$

where $\mathrm{D}$ denotes the up-shift (forward or right-shift) operator, $\mathrm{D}\, J_n = J_{n+1}$, and $\mathrm{I}$ is the identity operator.

The operator $\Delta$ takes the role of a spatial derivative on the shifted variables as many DDEs arise from discretization of a PDE in $1+1$ variables. Most, but not all, densities are polynomial in $\mathbf{u}_n$.

**2.11. Example**

The first three density-flux pairs [11] for (2) are

$$\rho_n^{(0)} = \ln(u_n), \qquad J_n^{(0)} = u_n + u_{n-1}, \tag{10}$$

$$\rho_n^{(1)} = u_n, \qquad J_n^{(1)} = u_n\, u_{n-1}, \tag{11}$$

$$\rho_n^{(2)} = \frac{1}{2} u_n^2 + u_n\, u_{n+1}, \qquad J_n^{(2)} = u_{n-1} u_n (u_n + u_{n+1}). \tag{12}$$

**2.12. Example**

The first four density-flux pairs [22] for (4) are

$$\rho_n^{(0)} = \ln(v_n), \qquad J_n^{(0)} = u_n, \tag{13}$$

$$\rho_n^{(1)} = u_n, \qquad J_n^{(1)} = v_{n-1}, \tag{14}$$

$$\rho_n^{(2)} = \frac{1}{2} u_n^2 + v_n, \qquad J_n^{(2)} = u_n v_{n-1}, \tag{15}$$

$$\rho_n^{(3)} = \frac{1}{3} u_n^3 + u_n (v_{n-1} + v_n), J_n^{(3)} = u_{n-1} u_n v_{n-1} + v_{n-1}^2. \tag{16}$$

The densities in (13)-(16) are uniform of ranks $0$ through 3, respectively. The corresponding fluxes are also uniform in rank with ranks $1$ through 4, respectively. In general, if in (8) rank $\rho_n = R$ then rank $J_n = R+1,$ since $w(\mathrm{D}_t)=1.$ The various pieces in (8) are uniform in rank. Since (8) holds on solutions of (1), the conservation law 'inherits' the dilation symmetry of (1).

Consult [22] for our algorithm to compute polynomial conserved densities and fluxes, where we use (4) to illustrate the steps. Non-polynomial densities (which are rare) can be computed by hand or with the method given in [8].

### 2.13. Definition

A vector function $\mathbf{G}(\mathbf{u}_n)$ is called a generalized (higher-order) symmetry of (1) if the infinitesimal transformation $\mathbf{u}_\mathrm{n} \to \mathbf{u}_\mathrm{n} + \varepsilon\, \mathbf{G}$ leaves (1) invariant up to order $\varepsilon.$ Consequently, $\mathbf{G}$ must satisfy [23]

$$\mathrm{D}_\mathrm{t}\mathbf{G} = \mathbf{F}'(\mathbf{u}_n)[\mathbf{G}] \tag{17}$$

on solutions of (1). $\mathbf{F}'(\mathbf{u}_n)[\mathbf{G}]$ is the Fréchet derivative of $\mathbf{F}$ in the direction of $\mathbf{G}.$

For the scalar case $(N=1),$ the Fréchet derivative in the direction of $G$ is computed as

$$F'(u_n)[G] = \frac{\partial}{\partial \varepsilon} F(u_n + \varepsilon\, G)|_{\varepsilon=0} = \sum_k \frac{\partial F}{\partial u_{n+k}} \mathrm{D}^k G, \tag{18}$$

which defines the Fréchet derivative operator

$$F'(u_n) = \sum_k \frac{\partial F}{\partial u_{n+k}} \mathrm{D}^k. \tag{19}$$

For the vector case with two components $u_n$ and $v_n,$ the Fréchet derivative operator is

$$\mathbf{F}'(\mathbf{u}_n) = \begin{pmatrix} \sum_k \frac{\partial F_1}{\partial u_{n+k}} \mathrm{D}^k & \sum_k \frac{\partial F_1}{\partial v_{n+k}} \mathrm{D}^k \\ \sum_k \frac{\partial F_2}{\partial u_{n+k}} \mathrm{D}^k & \sum_k \frac{\partial F_2}{\partial v_{n+k}} \mathrm{D}^k \end{pmatrix}. \tag{20}$$

Applied to $\mathbf{G} = (G_1, G_2)^\mathrm{T},$ where T is transpose, one gets

$$F_i'(\mathbf{u}_n)[\mathbf{G}] = \sum_k \frac{\partial F_i}{\partial u_{n+k}} \mathrm{D}^k G_1 + \sum_k \frac{\partial F_i}{\partial v_{n+k}} \mathrm{D}^k G_2, \quad i = 1,\ 2. \tag{21}$$

In (18) - (21), summation is over all positive and negative shifts (including the term without shift, i.e., $k=0$). For $k>0,$ $\mathrm{D}^\mathrm{k} = \mathrm{D} \circ \mathrm{D} \circ \ldots \circ \mathrm{D}$ ($k$ times). Similarly, for $k<0$ the down-shift operator $\mathrm{D}^{-1}$ is applied repeatedly. The generalization of (20) to $N$ components should be obvious.

### 2.14. Example

The first two symmetries [11] of (2) are

$$G^{(1)} = u_n(u_{n+1} - u_{n-1}), \tag{22}$$

$$G^{(2)} = u_n u_{n+1}(u_n + u_{n+1} + u_{n+2}) - u_{n-1}u_n(u_{n-2} + u_{n-1} + u_n). \quad (23)$$

These symmetries are uniform in rank (rank 2 and 3, respectively). Symmetries of ranks 0 and 1 are both zero.

### 2.15. Example

The first two non-trivial symmetries [24] of (4),

$$G^{(1)} = \begin{pmatrix} v_n - v_{n-1} \\ v_n(u_{n+1} - u_n) \end{pmatrix}, \quad (24)$$

$$G^{(2)} = \begin{pmatrix} v_n(u_n + u_{n+1}) - v_{n-1}(u_{n-1} + u_n) \\ v_n(u_{n+1}^2 - u_n^2 + v_{n+1} - v_{n-1}) \end{pmatrix}, \quad (25)$$

are uniform in rank. For example, rank $G_1^{(2)} = 3$ and rank $G_2^{(2)} = 4$. The symmetries of lower ranks are trivial.

An algorithm to compute polynomial generalized symmetries is described in detail in [24].

## 3. COMPUTATION OF RECURSION OPERATORS

### 3.1. Definition

A recursion operator $\Re$ connects symmetries

$$\mathbf{G}^{(j+s)} = \Re\, \mathbf{G}^{(j)}, \quad (26)$$

where $j = 1, 2, ...,$ and $s$ is the gap length. The symmetries are linked consecutively if $s = 1$. This happens in most, but not all, cases. For $N-$component systems, $\Re$ is an $N$ x $N$ matrix operator.

The defining equation for $\Re$ [6, 23] is

$$\mathrm{D_t}\Re + [\Re, \mathbf{F}'(\mathbf{u}_n)] = \frac{\partial \Re}{\partial t} + \Re'[\mathbf{F}] + \Re \circ \mathbf{F}'(\mathbf{u}_n) - \mathbf{F}'(\mathbf{u}_n) \circ \Re = 0, \quad (27)$$

where the bracket $[\,,\,]$ denotes the commutator of operators and $\circ$ the composition of operators. The operator $\mathbf{F}'(\mathbf{u}_n)$ was defined in (20). $\Re'[\mathbf{F}]$ is the Fréchet derivative of $\Re$ in the direction of $\mathbf{F}$. For the scalar case, the operator $\Re$ is often of the form

$$\Re = U(u_n)\mathrm{O}((\mathrm{D}-\mathrm{I})^{-1}, \mathrm{D}^{-1}, \mathrm{I}, \mathrm{D})V(u_n), \quad (28)$$

and in that case

$$\Re'[F] = \sum_k (\mathrm{D}^k F)\frac{\partial U}{\partial u_{n+k}}\mathrm{O}V + \sum_k U\,\mathrm{O}(\mathrm{D}^k F)\frac{\partial V}{\partial u_{n+k}}. \quad (29)$$

For the vector case and the examples under consideration, the elements of the $N$ x $N$ operator matrix $\Re$ are of the form $\Re_{ij} = U_{ij}(\mathbf{u}_n)\mathrm{O}_{ij}((\mathrm{D}-\mathrm{I})^{-1}, \mathrm{D}^{-1}, \mathrm{I}, \mathrm{D})V_{ij}(\mathbf{u}_n)$. Thus, for the two-component case [7]

$$\Re'[F]_{ij} = \sum_k (\mathrm{D}^k F_1)\frac{\partial U_{ij}}{\partial u_{n+k}} \mathrm{O}_{ij}\, V_{ij} + \sum_k (\mathrm{D}^k F_2)\frac{\partial U_{ij}}{\partial v_{n+k}} \mathrm{O}_{ij}\, V_{ij}$$
$$+ \sum_k U_{ij}\, \mathrm{O}_{ij} (\mathrm{D}^k F_1)\frac{\partial V_{ij}}{\partial u_{n+k}} + \sum_k U_{ij}\, \mathrm{O}_{ij} (\mathrm{D}^k F_2)\frac{\partial V_{ij}}{\partial v_{n+k}}. \tag{30}$$

### 3.2. Example

The KvM lattice (2) has recursion operator [7]

$$\begin{aligned}\Re &= u_n(\mathrm{I}+\mathrm{D})(u_n\mathrm{D} - \mathrm{D}^{-1}u_n)(\mathrm{D}-\mathrm{I})^{-1}\frac{1}{u_n}\mathrm{I} \\ &= u_n\mathrm{D}^{-1} + (u_n + u_{n+1})\mathrm{I} + u_n\mathrm{D} + u_n(u_{n+1} - u_{n-1})(\mathrm{D}-\mathrm{I})^{-1}\frac{1}{u_n}\mathrm{I}.\end{aligned} \tag{31}$$

### 3.3. Example

The Toda lattice (4) has recursion operator [7]

$$\Re = \begin{pmatrix} u_n\,\mathrm{I} & \mathrm{D}^{-1} + \mathrm{I} + (v_n - v_{n-1})(\mathrm{D}-\mathrm{I})^{-1}\frac{1}{v_n}\mathrm{I} \\ v_n\,\mathrm{I} + v_n\mathrm{D} & u_{n+1}\mathrm{I} + v_n(u_{n+1} - u_n)(\mathrm{D}-\mathrm{I})^{-1}\frac{1}{v_n}\mathrm{I} \end{pmatrix}. \tag{32}$$

### 3.4. Algorithm for computation of recursion operators

We will now construct the recursion operator (32) for (4). In this case all the terms in (27) are $2 \times 2$ matrix operators. The construction uses the following steps:

***Step 1 (Determine the rank of the recursion operator):*** The difference in rank of symmetries is used to compute the rank of the elements of the recursion operator.

Using (7), (24) and (25),

$$\operatorname{rank}\mathbf{G}^{(1)} = \begin{pmatrix} 2 \\ 3 \end{pmatrix}, \quad \operatorname{rank}\mathbf{G}^{(2)} = \begin{pmatrix} 3 \\ 4 \end{pmatrix}. \tag{33}$$

Assuming that $\Re\ \mathbf{G}^{(1)} = \mathbf{G}^{(2)}$, we use the formula

$$\operatorname{rank}\Re_{ij} = \operatorname{rank}\mathbf{G}_i^{(k+1)} - \operatorname{rank}\mathbf{G}_j^{(k)}, \tag{34}$$

to compute a rank matrix associated to the operator $\Re$

$$\operatorname{rank}\Re = \begin{pmatrix} 1 & 0 \\ 2 & 1 \end{pmatrix}. \tag{35}$$

***Step 2 (Determine the form of the recursion operator):*** $\Re = \Re_0 + \Re_1$ where $\Re_0$ is a sum of terms involving $\mathrm{D}^{-1}, \mathrm{I}$, and $\mathrm{D}$. The coefficients of these terms are admissible power combinations of $u_n, u_{n+1}, v_n$, and $v_{n-1}$ (which come from the terms on the right hand sides of (4)), so that all the terms have the correct rank. The maximum up-shift and down-shift operator that should be included can be determined by comparing two consecutive symmetries. Indeed, if the maximum up-shift in the first symmetry is $u_{n+p}$, and the maximum up-shift in the next symmetry is $u_{n+p+r}$, then the associated piece that goes into $\Re_0$ must have $\mathrm{D}, \mathrm{D}^2, ..., \mathrm{D}^{\mathrm{r}}$. The same line of reasoning determines the minimum down-shift operator to be included. So, in our example

$$\Re_0 = \begin{pmatrix} (\Re_0)_{11} & (\Re_0)_{12} \\ (\Re_0)_{21} & (\Re_0)_{22} \end{pmatrix}, \tag{36}$$

with

$$\begin{aligned} (\Re_0)_{11} &= (c_1 u_n + c_2 u_{n+1})\,\mathrm{I}, \\ (\Re_0)_{12} &= c_3 \mathrm{D}^{-1} + c_4\,\mathrm{I}, \\ (\Re_0)_{21} &= (c_5\, u_n^2 + c_6\, u_n u_{n+1} + c_7\, u_{n+1}^2 + c_8 v_{n-1} + c_9\, v_n)\,\mathrm{I} \\ &\quad + (c_{10}\, u_n^2 + c_{11} u_n u_{n+1} + c_{12}\, u_{n+1}^2 + c_{13} v_{n-1} + c_{14} v_n)\,\mathrm{D}, \\ (\Re_0)_{22} &= (c_{15} u_n + c_{16} u_{n+1})\,\mathrm{I}. \end{aligned} \tag{37}$$

As in the continuous case [10], $\Re_1$ is a linear combination (with constant coefficients $\tilde{c}_{jk}$ of sums of all suitable products of symmetries and covariants (Fréchet derivatives of conserved densities) sandwiching $(\mathrm{D} - \mathrm{I})^{-1}$. Hence,

$$\sum_j \sum_k \tilde{c}_{jk} \mathbf{G}^{(j)} (\mathrm{D} - \mathrm{I})^{-1} \otimes \rho_n^{(k)\prime}, \tag{38}$$

where $\otimes$ denotes the matrix outer product, defined as

$$\begin{pmatrix} G_1^{(j)} \\ G_2^{(j)} \end{pmatrix} (\mathrm{D}-\mathrm{I})^{-1} \otimes \begin{pmatrix} \rho_{n,1}^{(k)\prime} & \rho_{n,2}^{(k)\prime} \end{pmatrix} = \begin{pmatrix} G_1^{(j)} (\mathrm{D}-\mathrm{I})^{-1} \rho_{n,1}^{(k)\prime} & G_1^{(j)} (\mathrm{D}-\mathrm{I})^{-1} \rho_{n,2}^{(k)\prime} \\ G_2^{(j)} (\mathrm{D}-\mathrm{I})^{-1} \rho_{n,1}^{(k)\prime} & G_2^{(j)} (\mathrm{D}-\mathrm{I})^{-1} \rho_{n,2}^{(k)\prime} \end{pmatrix}. \tag{39}$$

Only the pair $(\mathbf{G}^{(1)}, \rho_n^{(0)\prime})$ can be used, otherwise the ranks in (35) would be exceeded. Using (13) and (21), we compute

$$\rho_n^{(0)\prime} = \begin{pmatrix} 0 & \dfrac{1}{v_n}\mathrm{I} \end{pmatrix}. \tag{40}$$

Therefore, using (38) and renaming $\tilde{c}_{10}$ to $c_{17}$,

$$\Re_1 = \begin{pmatrix} 0 & c_{17}(v_{n-1} - v_n)(\mathrm{D} - \mathrm{I})^{-1} \dfrac{1}{v_n}\mathrm{I} \\ 0 & c_{17} v_n (u_n - u_{n+1})(\mathrm{D} - \mathrm{I})^{-1} \dfrac{1}{v_n}\mathrm{I} \end{pmatrix}. \tag{41}$$

Adding (36) and (41), we obtain

$$\Re = \Re_0 + \Re_1 = \begin{pmatrix} \Re_{11} & \Re_{12} \\ \Re_{21} & \Re_{22} \end{pmatrix}. \tag{42}$$

***Step 3 (Determine the unknown coefficients):*** All the terms in (27) need to be computed. Referring to [7] for details, the result is:

$$\begin{aligned} &c_2 = c_5 = c_6 = c_7 = c_8 = c_{10} = c_{11} = c_{12} = c_{13} = c_{15} = 0, \\ &c_1 = c_3 = c_4 = c_9 = c_{14} = c_{16} = 1, \text{ and } c_{17} = -1. \end{aligned} \tag{43}$$

Substituting these constants into (42) finally gives

$$\Re = \begin{pmatrix} u_n \mathrm{I} & \mathrm{D}^{-1} + \mathrm{I} + (v_n - v_{n-1})(\mathrm{D}-\mathrm{I})^{-1} \frac{1}{v_n}\mathrm{I} \\ v_n \mathrm{I} + v_n \mathrm{D} & u_{n+1}\mathrm{I} + v_n(u_{n+1} - u_n)(\mathrm{D}-\mathrm{I})^{-1}\frac{1}{v_n}\mathrm{I} \end{pmatrix}. \tag{44}$$

One can readily verify that $\Re\,\mathbf{G}^{(1)} = \mathbf{G}^{(2)}$ with $\mathbf{G}^{(1)}$ in (24) and $\mathbf{G}^{(2)}$ in (25).

## 4. THE MATHEMATICA PACKAGE

To use the code, first load the *Mathematica* package **DDERecursionOperator.m** using the command

```
In[2]:= Get["DDERecursionOperator.m"];
```

Proceeding with the KvM lattice (2) as an example, call the function **DDERecursionOperator** (which is part of the package) :

```
In[3]:= DDERecursionOperator[{D[u[n,t],t]-(u[n,t]*(u[n+1,t]-u[n-1,t]))==0},
        {u},{n,t}]

Weight :: dilation : Dilation symmetry of the equation(s) is
{Weight[t]  -> -1, Weight[u]  -> 1}.

Out[3]= {{{DiscreteShift[#1,{n,-1}] u[n,t] + DiscreteShift[#1,{n,1}] u[n,t] + #1 (u[n,t]
+ u[1+n,t]) + Δ_n^-1[#1/u[n,t],{n,t}](-u[-1+n,t] u[n,t] + u[n,t] u[1+n,t]) &}}}
```

Here $\Delta_{\mathrm{n}}^{-1} = (\mathrm{D}-\mathrm{I})^{-1}$. The first part of the output (which we assign to R for later use) is indeed the recursion operator given in (31).

```
In[4]:= R = First[%];
```

Now using the first symmetry, generate the next symmetry by calling the function **GenerateSymmetries** (which is also part of the package):

```
In[5]:= firstsymmetry = {u[n,t](u[n+1,t]-u[n-1,t])};
In[6]:= GenerateSymmetries[R, firstsymmetry,1][[1]]
Out[6]= {-u[-2+n,t] u[-1+n,t] u[n,t] - u[-1+n,t]^2 u[n,t] - u[-1+n,t] u[n,t]^2
+ u[n,t]^2 u[1+n,t] + u[n,t] u[1+n,t]^2 + u[n,t] u[1+n,t] u[2+n,t]}
```

Evaluating the next five symmetries starting from the first one, can be done as follows:

In[7]:= TableForm[GenerateSymmetries[R, firstsymmetry, 5]]

Due to the length of the output we do not show this result here. The *Mathematica* function TableForm will nicely reformat the output in a tabular form. Our package is available at [15].

## 5. ADDITIONAL EXAMPLES

### 5.1. Ablowitz-Ladik (AL) Lattice

The AL lattice [19]

$$\begin{aligned} \dot{u}_n &= (u_{n+1} - 2u_n + u_{n-1}) + u_n v_n (u_{n+1} + u_{n-1}), \\ \dot{v}_n &= -(v_{n+1} - 2v_n + v_{n-1}) - u_n v_n (v_{n+1} + v_{n-1}), \end{aligned} \tag{45}$$

is an integrable discretization of the nonlinear Schrödinger (NLS) equation. The two recursion operators [7] computed by our package are:

$$\mathfrak{R}^{(1)} = \begin{pmatrix} \mathfrak{R}^{(1)}_{11} & \mathfrak{R}^{(1)}_{12} \\ \mathfrak{R}^{(1)}_{21} & \mathfrak{R}^{(1)}_{22} \end{pmatrix}, \tag{46}$$

with

$$\begin{aligned} \mathfrak{R}^{(1)}_{11} &= P_n \, \mathrm{D}^{-1} - u_n \, \Delta^{-1} v_{n+1} \mathrm{I} - u_{n-1} P_n \, \Delta^{-1} \frac{v_n}{P_n} \mathrm{I}, \\ \mathfrak{R}^{(1)}_{12} &= -u_n u_{n-1} \, \mathrm{I} - u_n \, \Delta^{-1} u_{n-1} \mathrm{I} - u_{n-1} P_n \, \Delta^{-1} \frac{u_n}{P_n} \mathrm{I}, \\ \mathfrak{R}^{(1)}_{21} &= v_n v_{n+1} \, \mathrm{I} + v_n \, \Delta^{-1} v_{n+1} \mathrm{I} + v_{n+1} P_n \, \Delta^{-1} \frac{v_n}{P_n} \mathrm{I}, \\ \mathfrak{R}^{(1)}_{22} &= (u_n v_{n+1} + u_{n-1} v_n) \mathrm{I} + P_n \, \mathrm{D} + v_n \, \Delta^{-1} u_{n-1} \mathrm{I} + v_{n+1} P_n \, \Delta^{-1} \frac{u_n}{P_n} \mathrm{I}, \end{aligned} \tag{47}$$

and

$$\mathfrak{R}^{(2)} = \begin{pmatrix} \mathfrak{R}^{(2)}_{11} & \mathfrak{R}^{(2)}_{12} \\ \mathfrak{R}^{(2)}_{21} & \mathfrak{R}^{(2)}_{22} \end{pmatrix}, \tag{48}$$

with

$$\begin{aligned} \mathfrak{R}^{(2)}_{11} &= P_n \, \mathrm{D} + u_n \, \Delta^{-1} v_{n-1} \mathrm{I} + (u_n v_{n-1} + u_{n+1} v_n) \mathrm{I} + u_{n+1} P_n \, \Delta^{-1} \frac{v_n}{P_n} \mathrm{I}, \\ \mathfrak{R}^{(2)}_{12} &= u_n u_{n+1} \, \mathrm{I} + u_n \, \Delta^{-1} u_{n+1} \mathrm{I} + u_{n+1} P_n \, \Delta^{-1} \frac{u_n}{P_n} \mathrm{I}, \\ \mathfrak{R}^{(2)}_{21} &= -v_n v_{n-1} \, \mathrm{I} - v_n \, \Delta^{-1} v_{n-1} \mathrm{I} - v_{n-1} P_n \, \Delta^{-1} \frac{v_n}{P_n} \mathrm{I}, \\ \mathfrak{R}^{(2)}_{22} &= P_n \, \mathrm{D}^{-1} - v_n \, \Delta^{-1} u_{n+1} \mathrm{I} - v_{n-1} P_n \, \Delta^{-1} \frac{u_n}{P_n} \mathrm{I}, \end{aligned} \tag{49}$$

where $P_n = 1 + u_n v_n$ and $\Delta = \mathrm{D} - \mathrm{I}$ . It can be shown that $\mathfrak{R}^{(1)} \circ \mathfrak{R}^{(2)} = \mathfrak{R}^{(2)} \circ \mathfrak{R}^{(1)} = \mathrm{I}$.

**5.2. Relativistic Toda (RT) Lattice**

The RT lattice [20] is given as

$$\dot{v}_n = v_n(u_{n-1} - u_n), \quad \dot{u}_n = u_n(u_{n-1} - u_{n+1} - v_{n+1} + v_n), \tag{50}$$

and the recursion operator found by our package coincides with the one in [20]:

$$\Re = \begin{pmatrix} v_n \mathrm{I} & v_n \mathrm{D}^{-1} + v_n \mathrm{I} + v_n(u_n - u_{n-1})(\mathrm{D}-\mathrm{I})^{-1}\frac{1}{u_n}\mathrm{I} \\ u_n \mathrm{D} + u_n \mathrm{I} & \begin{matrix} u_n\mathrm{D}^{-1} + u_n\mathrm{D} + (u_n + u_{n+1} + v_{n+1})\mathrm{I} \\ + u_n(u_{n+1} - u_{n-1} + v_{n+1} - v_n)(\mathrm{D}-\mathrm{I})^{-1}\frac{1}{u_n}\mathrm{I} \end{matrix} \end{pmatrix}. \tag{51}$$

## 6. CONCLUDING REMARKS

The existence of a recursion operator is a corner stone in establishing the complete integrability of nonlinear DDEs because the recursion operators allows one to compute an infinite sequence of generalized symmetries.

Therefore, we presented an algorithm to compute recursion operators of nonlinear DDEs with polynomial terms. The algorithm uses the scaling properties, conservation laws, and generalized symmetries of the DDE, but does not require the knowledge of the bi-Hamiltonian operators. The algorithm has been implemented in *Mathematica*, a leading computer algebra system. The package **DDERecursionOperators.m** uses **InvariantsSymmetries.m** to compute the conservation laws and higher-order symmetries of nonlinear DDEs.

The algorithm presented in this paper works for many nonlinear DDEs, including the Kac-van Moerbeke (Volterra), modified Volterra, and Ablowitz-Ladik lattices, as well as standard and relativistic Toda lattices. However, the algorithm does not allow one to compute recursion operators for lattices due to Blaszak-Marciniak and Belov-Chaltikian (see, e.g., [20] for references). An extension of the algorithm that would cover these lattices is under investigation.

**Acknowledgements-** This material is based upon work supported by the National Science Foundation (U.S.A.) under Grant No. CCF-0830783. J. A. Sanders, J.-P. Wang, M. Hickman and B. Deconinck are gratefully acknowledged for valuable discussions.